\documentclass[11pt]{article}
\usepackage{amsmath,amssymb,graphicx,cite,color,url,hyperref,setspace}
\usepackage[T1]{fontenc}
\def\beq{\begin{eqnarray}} \def\enq{\end{eqnarray}}

\def\lapp{\mathrel{\rlap{\raise.5ex\hbox{$<$}}{\lower.5ex\hbox{$\sim$}}}}
\def\gapp{\mathrel{\rlap{\raise.5ex\hbox{$>$}}{\lower.5ex\hbox{$\sim$}}}}
\begin{document} 

\begin{flushright} 
TIFR/TH/21-1
\end{flushright}

\begin{center} 
{\Large\bf A viable $\mathbf{U(1)}$ extended Standard Model with a 
massive $Z'$ \\ [2 mm] invoking the St\"{u}ckelberg mechanism} \\

\bigskip

{\large Radhika Vinze} \\ [1mm]
{\small Department of Physics, Univerity of Mumbai, \\
Vidyanagari, Santacruz (East), Mumbai 400098, India. \\ 
E-mail: \href{mailto:radhika.vinze@physics.mu.ac.in}
{\sf radhika.vinze@physics.mu.ac.in} }

{\large Sreerup Raychaudhuri} \\
{\small Department of Theoretical Physics, Tata Institute of Fundamental Research,  \\ 
Homi Bhabha Road, Mumbai 400\,005, India. \\ 
E-mail: \href{mailto:sreerup@theory.tifr.res.in}
{\sf sreerup@theory.tifr.res.in} }

\bigskip

\centerline{\sf\today}

\bigskip

{\large\bf ABSTRACT} 
\end{center} 

\vspace*{-0.3in} 

\setstretch{1.15} 
\small
\begin{quotation} 
{\noindent We make a careful re-examination of the possibility that, in 
a $U(1)$ extension of the Standard Model, the extra $Z'$ boson may 
acquire a mass from a St\"{u}ckelberg-type scalar. The model, when all 
issues of theoretical consistency are taken into account, contains 
several attractive new features, including a high degree of 
predictability.}
\end{quotation} 
\normalsize
\setstretch{1.25}

\centerline{PACS Nos:  12.15.-y, 12.60.-i, 14.80.-j}

\medskip

\hrule
\renewcommand{\theequation}{\arabic{section}.\arabic{equation}}
\setcounter{section}{0}
\setcounter{equation}{0}

\section{Introduction}

In recent times, it has become something of a clich\'e that physics 
beyond the Standard Model (SM) of elementary particles is like the Holy 
Grail of high energy physicists, proving equally elusive, and being 
characterised by many a myth \cite{SM}. And yet, every so often, an 
experimental result pops up which is not quite in conformity with the 
SM. The experience till now has been that such anomalies are either 
washed out with the collection of more statistics \cite{Anomalies}, or 
disappear after a re-evaluation of the SM prediction from a theoretical 
standpoint \cite{ThAnomalies}. Nevertheless, there appear to be a couple 
of such mismatches which have successfully resisted rapid extinction 
\cite{LHCb,Muon}, at least until the present juncture. Both relate to a 
seeming violation of lepton universality, one in the measurement of $B^0 
\to K^{*0}\ell^+\ell^-$ decay distributions \cite{LHCb}, and one in the 
measurement of the anomalous magnetic moment $\left(g-2\right)_\ell$, 
where $\ell = e,\mu$ \cite{Muon}. In either case, one of the more 
attractive 'beyond SM' (BSM) explanations of these processes is the 
presence of a massive $Z'$ vector boson with different couplings to the 
$e$ and the $\mu$ \cite{Zprime papers}. It is, therefore, natural to 
speculate if such a vector boson can be obtained in a simple bottom-up 
extension of the SM, and indeed, this is easily achieved if the 
$SU(2)\times U(1)_Y$ gauge symmetry of the SM is extended by an extra 
$U(1)_X$ symmetry to a $SU(2)\times U(1)_Y\times U(1)_X$ gauge symmetry 
\cite{Langacker}. The normal symmetry-breaking mechanism of the SM with 
a single scalar doublet will not, however, provide the extra $Z'$ boson 
with a mass, since it is protected by the extra $U(1)$ gauge symmetry. 
One must, therefore, also extend the {\it scalar} sector by adding an 
extra doublet or bi-doublet, which will break this symmetry and ensure 
that the $Z'$ acquires a mass \cite{U1breaking}. Such models, however, 
involve a mixing between the $Z$ and the $Z'$ bosons, which must be 
tuned very finely since the couplings of the $Z$ boson have been 
measured very precisely and seem to agree very closely with the SM 
predictions \cite{precisonZ}. Likewise, there will be mixing between the 
SM scalar doublet and the extra scalar multiplet(s), which will change 
the Higgs boson couplings accordingly. Since these are also constrained 
-- and are getting further constrained -- to be close to the SM values, 
this involves a whole set of further fine tuning \cite{precisonH}.

It is not that a phenomenologically viable model cannot be created using 
the above philosophy, and indeed, examples abound in the literature 
\cite{Zprime papers}, with a good deal of ingenuity having been expended 
in making these compatible with existing data. However, much of this 
effort can be avoided if we can devise a theory in which the $Z'$ boson 
acquires a mass {\it without} breaking the extra $U(1)$ symmetry. The 
advantage of such a theory would be that the presence of an unbroken 
symmetry would prevent a large number of operators from appearing in the 
Lagrangian, which otherwise appear through the symmetry-breaking. Such a 
model would, then be much more economical than the usual models, and 
consequently, far more predictive. It is in the pursuit of such a model 
that we turn to the idea of a St\"{u}ckelberg mechanism to generate the 
$Z'$ mass.

Many years before the application of spontaneous symmetry-breaking and 
the Higgs mechanism to electroweak theory, St\"{u}ckelberg had studied a 
$U(1)$ gauge theory and devised a gauge invariant way to give mass to 
the gauge boson. St\"{u}ckelberg's model \cite{Stuck} involved adding an 
extra `gauge scalar' to the Lagrangian, and assigning to it a specific 
gauge transformation which would keep the action invariant even when the 
gauge boson mass terms were included. In the original model, however, 
the scalar and the gauge boson became mass-degenerate, and the obvious 
absence of such accompanying scalar particles led to the early demise of 
the idea. The concept of spontaneous symmetry-breaking and the Higgs 
mechanism which came up subsequently, proved to be far more successful 
\cite{Higgs}. That the electroweak interactions can be accurately 
described by the Higgs scalar-based model is, of course, no longer in 
doubt, following the discovery of the Higgs boson and the close 
correlation observed between particles masses and couplings as predicted 
in the SM \cite{precisonH}. Nevertheless, the St\"{u}ckelberg mechanism 
remains an attractive idea for generating masses of gauge bosons, and it 
may well be a path which Nature chooses {\it in addition to} the 
spontaneous symmetry-breaking route. Obviously, this will not form any 
part of the SM, but it can still prove useful in extensions of the SM 
which envisage the existence of extra gauge bosons, such as, for 
example, the massive $Z'$ in a $U(1)$ extension of the SM discussed 
above.

In this article, therefore, we explore the possibility that the mass of 
the $Z'$ in a $U(1)$ extended SM can arise from a St\"{u}ckelberg mechanism. 
This requires the further extension of the model by one St\"{u}ckelberg 
scalar, but no extension of the Higgs sector. This model, as will be shown, 
is very economical with parameters. Such ideas have been studied before 
\cite{PranNath}, but our work makes a thorough investigation of the model, 
which turns out to be more restrictive when all considerations of internal 
consistency are taken into account. We have also worked out the mass spectrum 
and Feynman rules relevant to this model. In a subsequent work, we shall be 
exploring the phenomenological implications and possible experimental 
signatures which could be used to verify these ideas \cite{selves}. 

This article is organised as follows. The next two sections serve to set 
the notation and also provide the reader with a quick introduction so 
that this article may be read independently without consulting the 
references. Thus, in Section 2, we briefly introduce the original idea 
of St\"{u}ckelberg and show why the model was not viable. In Section 3, 
we describe a $U(1)$ extension of the SM and show why the $Z'$ boson 
must remain massless unless an extra mechanism is introduced. The core 
of our work is described in Section 4, where we show how the 
introduction of a St\"{u}ckelberg scalar can give a mass to the $Z'$. 
Section 5 is devoted to a study of the model parameter space and the new 
Feynman vertices for the theory. Some concluding remarks and caveats are 
relegated to the final section.

\section{The $\mathbf{U(1)}$ St\"{u}ckelberg model}
\setcounter{equation}{0}

In this section we briefly review the original idea of St\"{u}ckelberg 
\cite{Stuck}, and apply it to a minimal model with a $U(1)$ gauge 
symmetry. As is well known, if we consider a model with a single fermion 
field $\psi(x)$ with a local $U(1)$ gauge transformation
\begin{equation}
\psi(x) \to \psi'(x) =  \exp[-ig\theta(x)] \ \psi(x)
\label{eqn:gauge-trans-U1}
\end{equation}
then, a Lagrangian which is invariant under this local gauge invariance 
(\ref{eqn:gauge-trans-U1}) will have the form
\begin{equation}
{\cal L}_{\rm U(1)} = i \bar{\psi}  \gamma^\mu D_\mu \psi  - m  \bar{\psi}  \psi 
- \frac{1}{4} F^{\mu\nu}  F_{\mu\nu}  
\label{eqn:lagrangian-U1}
\end{equation}
where $D_\mu = \partial_\mu - igA_\mu $ and the gauge field $A_\mu(x)$ 
has the transformation
\begin{equation}
A_\mu(x) \to A'_\mu(x) = A_\mu(x) + \partial_\mu \theta(x)
\label{eqn:trans-U1}
\end{equation}
under which $F_{\mu\nu} = \partial_\mu A_\nu - \partial_\nu A_\mu$ is 
invariant. The gauge symmetry, however, forbids the writing of a mass 
term for the gauge boson, which would be
\begin{equation}
{\cal L}_M = \frac{1}{2}  M ^2 A^\mu  A_\mu 
\label{eqn:mass-U1}
\end{equation}
and this is the oft-quoted reason for the photon to be massless, since 
the above Lagrangian is simply the QED Lagrangian.

The common way to generate a gauge boson mass is by introducing into the 
model a self-interacting tachyonic Higgs scalar which induces 
spontaneous breakdown of the gauge symmetry at low energies, allowing a 
gauge boson mass to develop, as well as acquiring a real mass for 
itself. This is, of course, the famous 
Brout-Englert-Higgs-Guralnik-Hagen-Kibble (BEHGHK) mechanism 
\cite{Higgs} and the massive scalar is a Higgs boson.

St\"{u}ckelberg's idea -- which pre-dated the BEHGHK mechanism -- was to 
introduce into the model a scalar field $\sigma(x)$, which would have a 
gauge transformation
\begin{equation}
\sigma(x) \to \sigma'(x) = \sigma(x) -  M  \theta(x)
\label{eqn:scalar-U1}
\end{equation}
which would render the construct
\begin{equation}
\Gamma_\mu = A_\mu   + \frac{1}{ M } \partial_\mu \sigma 
\label{eqn:Stu-invariant1-U1}
\end{equation}
gauge invariant if the gauge field $A_\mu(x)$ transforms as in 
Eq.~(\ref{eqn:trans-U1}). The mass term can then be rewritten as
\begin{equation}
{\cal L}_{\rm M} = \frac{1}{2}  M ^2 \Gamma_\mu \Gamma^\mu 
\label{eqn:mass-Stuck-U1}
\end{equation}
This expands to give a gauge boson mass term as well as a kinetic term 
for the scalar $ \sigma(x)$ field. However, there is no mass term for 
the $\sigma(x)$ and there is also an extra bilinear term $ M A^\mu 
\partial_\mu \sigma$, which cannot be physically interpreted. This led 
to the demise of the original St\"{u}ckelberg model.

However, there is another construct that can be made, and this is
\begin{equation}
\Delta = \sigma  - \frac{1}{ M } \partial_\mu A^\mu 
\label{eqn:Stu-invariant2-U1}
\end{equation}
which is gauge invariant so long as we stay within the family of 
harmonic gauges, i.e. satisfying $\Box \theta(x) = 0$. If we accept this 
restricted definition of gauge invariance, we can add a term
\begin{equation}
{\cal L}_M^\sigma = -\frac{1}{2}  M ^2 \Delta^2 
\label{eqn:masspar-Stuck-U1}
\end{equation}
to the Lagrangian, which provides a mass term for the scalar, as well as 
a bilinear term $M \sigma \partial_\mu A^\mu$, and a gauge fixing term
$$
-\frac{1}{2} \left\{ \partial_\mu A^\mu  \right\}^2 
\label{eqn:fixing-Stuck-U1}
$$
It is clear that the two bilinear terms combine to give a total 
derivative which can be dropped from the Lagrangian, and what we get 
finally is
\begin{equation}
{\cal L}_{\rm S} = i \bar{\psi} \gamma^\mu D_\mu \psi  - m  \bar{\psi}  \psi  
+  \frac{1}{2} \partial^\mu \sigma  \partial_\mu \sigma  -  \frac{1}{2} M ^2 \sigma^2 
 - \frac{1}{4} F^{\mu\nu}  F_{\mu\nu}  + \frac{1}{2}  M ^2 A^\mu  A_\mu 
- \frac{1}{2} \left( \partial_\mu A^\mu  \right)^2  
\label{eqn:Stuckelberg-U1}
\end{equation}
This is a very nice Lagrangian indeed, for it does not break the gauge 
symmetry and yet has a massive gauge boson, as well as a gauge fixing 
term with $\xi = 1$, indicative of the Feynman gauge (which in QED has 
to be put in by hand). However, the problem with this model is that the 
scalar $\sigma$ must be mass-degenerate with the gauge boson $A_\mu$ for 
the bilinear terms to combine, and this would make it non-viable as a 
model for any kind of theory of the weak interactions. As a result, 
despite its considerable elegance, the St\"{u}ckelberg theory was 
abandoned and has remained a curiosity for the past several decades.

\section{A $\mathbf{U(1)_X}$ extension of the Standard Model}
\setcounter{equation}{0}

For the moment, we omit fermions and describe the scalar and gauge 
sector of the SM, which is invariant under $SU(2)_L\times U(1)_Y$ gauge 
transformations,
\begin{equation}
\Phi (x) \to \Phi'(x) =  \exp \left(-ig \mathbb{T}_a \theta_a 
- \frac{i}{2} g'_Y Y_\Phi \theta' \mathbb{I} \right) \ \Phi (x)
\label{eqn:gauge-SM}
\end{equation}
with $\mathbb{T}_a = \frac{1}{2}\sigma_a \ (a=1,2,3)$ being the $SU(2)$ 
generators and $\theta_a, \theta'$ being the parameters of the gauge 
transformation, while $Y_\Phi$ is the weak hypercharge of the $\Phi$ 
field. The SM Lagrangian is
\begin{equation}
{\cal L}_{\rm SM} = \left(\mathbb{D}^\mu \Phi\right)^\dagger \mathbb{D}_\mu \Phi  
-\frac{1}{8} {\rm Tr}(\mathbb{W}_{\mu\nu}\mathbb{W}^{\mu\nu}) 
- \frac{1}{4} B_{\mu\nu} B^{\mu\nu} - V(\Phi)
\label{eqn:lagrangian-SM}
\end{equation}
where, as usual,
\begin{eqnarray}
\mathbb{D}_\mu & = & \mathbb{I}\partial_\mu - ig \mathbb{T}_a W_\mu^a 
-\frac{i}{2} g'_Y Y_\Phi B_\mu \mathbb{I}  
\hspace*{1.0in}
\mathbb{W}_{\mu\nu} = \frac{1}{ig} \left[\mathbb{D}_\mu,\mathbb{D}_\nu \right] 
\\
V(\Phi) & = & -\mu^2 \Phi^\dagger \Phi + \lambda \left(\Phi^\dagger \Phi\right)^2
\hspace*{1.45in}
B_{\mu\nu} = \partial_\mu B_\nu - \partial_\nu B_\mu
\nonumber 
\label{eqn:covariants-SM}
\end{eqnarray}
in terms of vector gauge fields $W_\mu^a \ (a = 1,2,3)$ and $B_\mu$, and the doublet of 
scalar fields
\begin{equation}
\Phi = \left( \begin{array}{c} \varphi^+ \\ 
                               \frac{1}{\sqrt{2}}\left(v + h^0 +i\varphi^0\right)
              \end{array} \right)
\label{eqn:higgs-SM}
\end{equation}
where $v^2 = \mu^2/2\lambda$. As is well known, the $SU(2)_L\times 
U(1)_Y$ gauge symmetry is spontaneously broken by the $v$ parameter in 
Eq.~(\ref{eqn:higgs-SM}), leading to masses
\begin{equation}
M_W = \frac{1}{2}gv  \hspace*{1.0in} 
M_Z = \frac{1}{2}gv \sec\theta_W \hspace*{1.0in} 
M_A = 0
\label{eqn:masses-SM}
\end{equation}
for the physical gauge bosons
\begin{equation}
W_\mu^\pm = \frac{1}{\sqrt{2}} \left(W_\mu^1 \mp iW_\mu^2\right)  \hspace*{0.5in} 
Z_\mu = W_\mu^3 \cos \theta_W - B_\mu \sin \theta_W \hspace*{0.5in} 
A_\mu = W_\mu^3 \sin \theta_W + B_\mu \cos \theta_W
\label{eqn:gaugebosons-SM}
\end{equation}
with $\tan \theta_W = g'_Y Y_\Phi/g$, and a scalar mass $M_h = 
\sqrt{2}\mu$, while the $\varphi^\pm$, $\varphi^0$ remain massless and 
can indeed be absorbed into the definitions of the gauge bosons by a 
judicious choice of the initial gauge (unitary gauge). In addition, when 
this theory is quantised, it will be necessary to add gauge fixing and 
Fadeev-Popov ghost terms to the Lagrangian, which are omitted here for 
the sake of brevity.

We now consider the modification of the above theory where the gauge 
symmetry is extended to $SU(2)_L\times U(1)_Y\times U(1)_X$, i.e. by an 
extra 'weak hypercharge' $X_\Phi$ in addition to the weak hypercharge 
$Y_\Phi$. The corresponding gauge transformation is
\begin{equation}
\Phi (x) \to \Phi'(x) =  \exp \left(-ig \mathbb{T}_a \theta_a 
- \frac{i}{2} g'_Y Y_\Phi \theta'_Y \mathbb{I} 
-  \frac{i}{2} g'_X X_\Phi \theta'_X \mathbb{I} \right) \ \Phi (x)
\label{eqn:gauge-SMX}
\end{equation}
and the Lagrangian becomes
\begin{equation}
{\cal L}_{\rm SMX} = \left(\mathbb{\bar{D}}^\mu \Phi\right)^\dagger \mathbb{\bar{D}}_\mu \Phi  
-\frac{1}{8} {\rm Tr}(\mathbb{W}_{\mu\nu}\mathbb{W}^{\mu\nu}) 
- \frac{1}{4} B_{\mu\nu} B^{\mu\nu} - \frac{1}{4} C_{\mu\nu} C^{\mu\nu} - V(\Phi)
\label{eqn:lagrangian-SMX}
\end{equation}
where, now
\begin{eqnarray}
\mathbb{\bar{D}}_\mu = \mathbb{I}\partial_\mu - ig \mathbb{T}_a W_\mu^a 
-\frac{i}{2} g'_Y Y_\Phi B_\mu \mathbb{I} -\frac{i}{2} g'_X X_\Phi C_\mu \mathbb{I} 
\hspace*{0.5in}
\mathbb{W}_{\mu\nu} & = & \frac{1}{ig} \left[\mathbb{D}_\mu,\mathbb{D}_\nu \right] 
\\
V(\Phi) = -\mu^2 \Phi^\dagger \Phi + \lambda \left(\Phi^\dagger \Phi\right)^2
\hspace*{1.6in}
B_{\mu\nu} & = & \partial_\mu B_\nu - \partial_\nu B_\mu
\nonumber \\
C_{\mu\nu} & = & \partial_\mu C_\nu - \partial_\nu C_\mu
\nonumber
\label{eqn:covariants-SMX}
\end{eqnarray} 
with a new gauge field $C_\mu$ with a new coupling constant $g'_X$, and 
with a new 'weak hypercharge' $X_\Phi$ of the scalar doublet. All other 
symbols have the same meanings as before.

As in the SM, the mass terms arise from the seagull terms
\begin{eqnarray}
{\cal L}_{\rm M} &=& \left[ 
\left(- ig \mathbb{T}_a W^{\mu a} -\frac{i}{2} g'_Y Y_\Phi B^\mu \mathbb{I} 
-\frac{i}{2} g'_X X_\Phi C^\mu \mathbb{I}\right) \langle\Phi\rangle \right]^\dagger 
\nonumber \\
& \times & \left[ \left(- ig \mathbb{T}_b W_\mu^b -\frac{i}{2} g'_Y Y_\Phi B_\mu \mathbb{I}
-\frac{i}{2} g'_X X_\Phi C_\mu \mathbb{I} \right) \langle\Phi\rangle \right] 
\label{eqn:seagull-SMX}
\end{eqnarray}
where
\begin{equation}
\langle\Phi\rangle = \left( \begin{array}{c} 0 \\ 
                               v/\sqrt{2}
              \end{array} \right)
\label{eqn:vev-SM}
\end{equation}
which leads to
\begin{equation}
\begin{array}{r}
{\cal L}_{\rm M} = M_W^2 W^{+\mu}W_\mu^- + \frac{1}{2} 
\left( \begin{array}{ccc} W^{3\mu} & B^\mu & C^\mu \end{array} \right)
\mathbb{M} \\ [1mm] \\ [1mm]
\end{array} 
\left( \begin{array}{c} W^3_\mu \\ B_\mu \\ C_\mu \end{array} \right)
\label{eqn:masses-SMX}
\end{equation}
where the mass matrix $\mathbb{M}$ is
\begin{equation}
\mathbb{M} = \frac{v^2}{4} \left( \begin{array}{ccc}
g^2     & gg'_Y Y_\Phi      & gg'_X X_\Phi      \\
gg'_Y Y_\Phi &(g'_Y Y_\Phi)^2    & g'_X g'_Y X_\Phi Y_\Phi \\
gg'_X X_\Phi & g'_X g'_Y X_\Phi Y_\Phi &(g'_X X_\Phi)^2 
\end{array} \right)
\label{eqn:matrix-SMX}
\end{equation}
One eigenvalue of this matrix is $\frac{1}{4} v^2 \left(g^2 + g^{\prime 
2}_Y Y_\Phi^2 + g^{\prime 2}_X X_\Phi^2 \right)$ and the other two are 
$0,0$. If we identify the corresponding mixed neutral boson states as 
the $Z$ boson, the photon $A$ and a new $Z'$ boson, respectively, then 
the mass of the $Z$ boson can be identified as
\begin{equation}
M_Z^2 = \frac{1}{4} v^2 \left(g^2 + g^{\prime 2}_Y Y_\Phi^2 + g^{\prime 2}_X X_\Phi^2 \right)
\label{eqn:Zmass-SMX}
\end{equation} 
while $M_\gamma = M_{Z'} = 0$. While the photon should indeed be 
massless, as in the Standard Model, a massless $Z'$ would obviously have 
phenomenological consequences which would have been detected long ago. 
Hence arises the urgency that the $Z'$ should acquire a mass.

In this model, of course, the zero mass of the $Z'$ boson may be 
directly traced to the extra $U(1)_X$ symmetry, as a result of which, 
the symmetry-breaking pattern in this model is
$$
SU(2)_L \times U(1)_Y \times U(1)_X \rightarrow  U(1)_{\rm em} \times U(1)_{Z'}
$$
with only three out of the five symmetry-generators being broken. To 
break the additional symmetry, it is usual to introduce an additional 
scalar which develops a vacuum expectation value of its own and breaks 
the residual $U(1)_{\rm em} \times U(1)_{Z'}$ symmetry to $U(1)_{\rm 
em}$ alone. As stated in the Introduction, there is nothing wrong in 
such an approach, since, after all, one cannot strain at the gnat of 
this additional symmetry-breaking after having swallowed the camel of 
the Higgs-sector symmetry-breaking. Nevertheless, breaking a symmetry 
always permits the inclusion of symmetry-breaking interactions with an 
attendant proliferation of undetermined parameters. These have then to 
be tuned for internal consistency and compatibility with experimental 
data. It is not our purpose, in this article, to critique the 
symmetry-breaking approach to obtain a massive $Z'$ boson, but simply to 
explore an alternative idea, viz. that the $U(1)_{\rm em} \times 
U(1)_{Z'}$ symmetry remains unbroken, but the $Z'$ acquires mass through 
a St\"uckelberg mechanism, i.e. through the addition of a 'gauge scalar' 
rather than a symmetry-breaking scalar as described above. We shall see 
that the actual number of parameters in this model will indeed be highly 
constrained, as expected from the existence of an unbroken symmetry. It 
will however, retain the necessary flexibility to provide 
generation-dependent couplings for the $Z'$ boson, which was the 
starting point of our argument.

At this juncture, it is necessary to mention that the idea of generating 
a $Z'$ mass through a St\"{u}ckelberg mechanism is not new. It has been 
explored in Refs. \cite{PranNath} with a fair degree of thoroughness. 
However, we have revisited the basic idea, imposing some extra 
consistency conditions and thereby obtaining a more restrictive theory. 
Our work, therefore, is intended to complete, rather than refute, 
earlier works on this interesting question.

\section{St\"{u}ckelberg masses in the $\mathbf{SU(2)_L\times U(1)_Y 
\times U(1)_X}$ model}
\setcounter{equation}{0}

In the model described in the previous section, the $B_\mu$ and $C_\mu$ 
gauge fields undergo the usual $U(1)$ gauge transformations, 
independently of the $W_\mu^a$, viz.
\begin{equation}
B_\mu \to B'_\mu = B_\mu + \partial_\mu \theta'_Y
\qquad\qquad
C_\mu \to C'_\mu = C_\mu + \partial_\mu \theta'_X
\label{eqn:abelian-SMXStuck}
\end{equation} 
under which the Lagrangian ${\cal L}_{\rm SMX}$ remains invariant. We 
now add a St\"{u}ckelberg scalar $\sigma_0$ which transforms under the 
same pair of gauge transformations as
\begin{equation}
\sigma_0 \to \sigma'_0 = \sigma_0 - M_Y \theta'_Y - M_X \theta'_X
\label{eqn:scalar-SMXStuck}
\end{equation}
in terms of the mass-dimension parameters $M_X, M_Y$. We can now 
construct the gauge invariant expressions
\begin{equation}
\Gamma_\mu = \partial_\mu \sigma_0 + M_X B_\mu + M_Y C_\mu 
\label{eqn:Stu-invariant1-SMXStuk}
\end{equation}
and
\begin{equation}
\Delta = \sigma_0 - \frac{1}{M_Y} \partial_\mu B^\mu - \frac{1}{M_X} \partial_\mu C^\mu 
\label{eqn:Stu-invariant2-SMXStuk}
\end{equation}
where, as in the Abelian case, we restrict the gauge invariance to the 
case of harmonic gauges satisfying $(\Box + M_Y^2) \theta'_Y = (\Box + 
M_X^2) \theta'_X = 0$. In terms of these we can now construct the 
St\"{u}ckelberg part of the Lagrangian as
\begin{equation}
{\cal L}_{\rm S} = 
\frac{1}{2M^2} \left( M^2 + \lambda'_1 \Phi^\dagger \Phi \right) \Gamma_\mu \Gamma^\mu 
- \frac{1}{2} \left( M^2 + \lambda'_2 \Phi^\dagger \Phi \right) \Delta^2  
\label{eqn:Stu-lagrangian-SMXStuk}
\end{equation}
where $M$ is a mass parameter, and $\lambda'_1$ and $\lambda'_2$ are new 
dimensionless coupling constants (not to be confused with the quartic 
self-coupling $\lambda$ in the embedded $V(\Phi)$). The complete 
Lagrangian is then obtained by combining those in 
Eqs.~(\ref{eqn:lagrangian-SMX}) and (\ref{eqn:Stu-lagrangian-SMXStuk}) 
as
\begin{equation}
{\cal L}_{\rm SMX}^{(\sigma)} = {\cal L}_{\rm SMX} + {\cal L}_{\rm S}
\label{eqn:model-SMXStuck}
\end{equation}
Apart from the St\"{u}ckelberg masses, further mass terms and bilinears 
will be generated by spontaneous symmetry-breaking in the Higgs sector, 
which will require the replacement of $\Phi^\dagger \Phi =v^2/2$ in the 
above equation. Expanding the terms which arise thereby, we obtain a 
free scalar Lagrangian
\begin{equation}
{\cal L}_\sigma = \frac{1}{2} \left( 1 + \frac{\lambda'_1 v^2}{2M^2} \right) 
 \partial_\mu \sigma_0 \, \partial^\mu \sigma_0
 -  \frac{1}{2} M^2 \left( 1 + \frac{\lambda'_2 v^2}{2M^2} \right) \sigma_0^2 
\label{eqn:freescalar-SMXStuck}
\end{equation} 
To obtain the proper normalisation for the kinetic term, it is necessary 
to renormalise the scalar $\sigma_0$, writing
\begin{equation}
\sigma_0 (x) = \frac{\sigma(x)}{\sqrt{Z_\sigma}}
\label{eqn:renorm-SMXStuck}
\end{equation}
where
\begin{equation}
Z_\sigma = 1 + \frac{\lambda'_1 v^2}{2M^2}
\label{eqn:renormfac-SMXStuck}
\end{equation}
In terms of the renormalised scalar defined in 
Eq.~(\ref{eqn:renorm-SMXStuck}), we can now rewrite the 
Eq.~(\ref{eqn:freescalar-SMXStuck}) in the form
\begin{equation}
{\cal L}_\sigma = \frac{1}{2}  \partial_\mu \sigma \, \partial^\mu \sigma
 -  \frac{1}{2} M^2_\sigma \sigma^2 
 \label{eqn:renorm-freescalar-SMXStuck}
\end{equation}
where
\begin{equation}
M_\sigma =  M \sqrt{\left(1 + \frac{\lambda'_2 v^2}{2M^2}\right)
\left(1 + \frac{\lambda'_1 v^2}{2M^2}\right)^{-1}} 
\label{eqn:scalarmass-SMXStuck}
\end{equation}
is the mass of the physical scalar.

The bilinear terms involving the $\sigma$ and the $B_\mu, C_\mu$ fields 
will take the form
\begin{eqnarray}
{\cal L}_{\rm bil} & = & 
  \left( 1 + \frac{\lambda'_1 v^2}{2M^2} \right) M_Y \ \partial_\mu \sigma_0 \, B^\mu 
+ \left( 1 + \frac{\lambda'_2 v^2}{2M^2} \right) \frac{M^2}{M_Y} \ \sigma_0 \, \partial_\mu B^\mu 
\nonumber \\
& + & \left( 1 + \frac{\lambda'_1 v^2}{2M^2} \right) M_X \ \partial_\mu \sigma_0 \, C^\mu 
+ \left( 1 + \frac{\lambda'_2 v^2}{2M^2} \right) \frac{M^2}{M_X} \ \sigma_0 \, \partial_\mu C^\mu
\label{eqn:bilinears-SMXStuck}
\end{eqnarray}
These will reduce to a pair of total derivatives and can be dropped from 
the Lagrangian -- as in the $U(1)$ case -- provided the coefficients 
satisfy
\begin{equation}
\left( 1 + \frac{\lambda'_1 v^2}{2M^2} \right) M_Y 
= \left( 1 + \frac{\lambda'_2 v^2}{2M^2} \right) \frac{M^2}{M_Y}
\label{eqn:bil-coeffY-SMXStuck}
\end{equation}
and
\begin{equation}
\left( 1 + \frac{\lambda'_1 v^2}{2M^2} \right) M_X 
= \left( 1 + \frac{\lambda'_2 v^2}{2M^2} \right) \frac{M^2}{M_X}
\label{eqn:bil-coeffX-SMXStuck}
\end{equation}
Imposing this constraint, it follows from 
Eqs.~(\ref{eqn:bil-coeffX-SMXStuck}) and (\ref{eqn:bil-coeffY-SMXStuck}) 
that
\begin{equation}
M_X^2 = M_Y^2
\label{eqn:MX-MY-SMXStuck}
\end{equation}
i.e. we must take $M_X = M_Y$ for cancellation of the bilinears, as well 
as a relation (\ref{eqn:bil-coeffY-SMXStuck}) between the parameters 
$M_Y$, $M$, $\lambda'_1$ and $ \lambda'_2$. It may be noted in passing 
that we could equally well have taken $M_X = -M_Y$. However, this merely 
amounts to a redefinition of the gauge parameter $\theta'_X$ and cannot 
change any of the physical results, since the $U(1)_X$ gauge symmetry 
remains unbroken.

In addition to the bilinear terms, expansion of the right side of 
Eq.~(\ref{eqn:Stu-lagrangian-SMXStuk}) leads to gauge-fixing terms
\begin{equation}
{\cal L}_{\rm gf} = -\frac{M^2}{2M_Y^2} \left( 1 + \frac{\lambda'_2 v^2}{2M^2} \right) 
\left[ \left( \partial_\mu B^\mu \right)^2  +  \left( \partial_\mu C^\mu \right)^2 
+ 2 \, \partial_\mu B^\mu \, \partial_\nu C^\nu \right]
\label{eqn:gauge-bilinears-SMXStuck}
\end{equation}
The last (bilinear) term on the right can be removed by redefining
\begin{equation}
\left( \!\!\! \begin{array}{c} B_\mu \\ C_\mu \end{array} \!\!\! \right) = 
\left( \!\!\! \begin{array}{cc} \frac{1}{\sqrt{2}} & \frac{1}{\sqrt{2}} \\ 
\frac{1}{\sqrt{2}} & -\frac{1}{\sqrt{2}} \end{array} \!\!\! \right)
\left( \!\!\! 
\begin{array}{c} \widetilde{B}_\mu \\ \widetilde{C}_\mu \end{array} \!\!\! \right)
\end{equation}
which leads to
\begin{equation}
{\cal L}_{\rm gf} = -\frac{1}{2\xi_B} \left( \partial_\mu \widetilde{B}^\mu \right)^2
-\frac{1}{2\xi_C} \left( \partial_\mu \widetilde{C}^\mu \right)^2
\label{eqn:fixing-SMXStuck}
\end{equation}
where
\begin{equation}
\xi_B = \frac{2M_Y^2}{M^2} \left( 1 + \frac{\lambda'_2 v^2}{2M^2} \right)^{-1}
\qquad\qquad \xi_C \to \infty
\label{eqn:fixing-param-SMXStuck}
\end{equation}
i.e. we have a specific gauge choice for the $\widetilde{B}_\mu$ field 
(as in the $U(1)$ theory) and the unitary gauge for the 
$\widetilde{C}_\mu$ field (or any other choice, e.g. the Feynman gauge, 
which must be put in by hand). For the remaining part of this 
discussion, these gauge choices will be assumed \footnote{It may be 
noted that we have arrived at the gauge choices by demanding the 
disappearance of bilinear terms in the Lagrangian. This is completely 
equivalent to the procedure in Ref.~\cite{Kuzmin}, where the gauge 
fixing terms are chosen so as to induce cancellation of the bilinears.}.

We are now in a position to write down the mass terms for the gauge 
bosons. Taking the constraints in Eqs.~(\ref{eqn:bil-coeffY-SMXStuck}) 
and (\ref{eqn:MX-MY-SMXStuck}) into account, we can now work out the 
mass terms as
\begin{equation}
\begin{array}{r} {\cal L}_{\rm M} = M_W^2 W^{+\mu}W_\mu^- + \frac{1}{2} 
\left( \begin{array}{ccc} W^{3\mu} & \widetilde{B}^\mu & \widetilde{C}^\mu \end{array} \right)
\mathbb{M} \\ [1mm] \\ [1mm] \end{array}
\left( \begin{array}{c} W^3_\mu \\ \widetilde{B}_\mu \\ \widetilde{C}_\mu \end{array} \right)
\label{eqn:masses-SMXStuk}
\end{equation}
where the mass matrix $\mathbb{M}$ is now
\begin{equation}
\mathbb{M} = \frac{M_W^2}{4} \left( \begin{array}{ccc}
2 & -\sqrt{2} a_\Phi & -\sqrt{2} b_\Phi \\
-\sqrt{2} a_\Phi & a_\Phi^2 + 4\mu^2 & a_\Phi b_\Phi \\
-\sqrt{2} b_\Phi & a_\Phi b_\Phi & b_\Phi^2 \end{array} \right)
\label{eqn:matrix-SMXStuk}
\end{equation}
where $M_W = \frac{1}{2}gv$, as usual, and
\begin{equation}
a_\Phi = \frac{1}{g} \left( g'_Y Y_\Phi + g'_X X_\Phi \right)
\qquad\qquad
b_\Phi = \frac{1}{g} \left( g'_Y Y_\Phi - g'_X X_\Phi \right)
\label{eqn:a-b-SMXStuk}
\end{equation}
while
\begin{equation}
\mu^2 = Z_\sigma \frac{M_\sigma^2}{M_W^2}
\label{eqn:mu-SMXStuk}
\end{equation}
It is easy to check that this matrix has eigenvalues which lead to real 
and non-negative gauge boson masses, for all values of $a_\Phi$, 
$b_\Phi$ and $\mu$, and that one of the eigenvalues is always zero, 
which can be identified with the photon mass. The non-zero eigenvalues 
are
\begin{equation}
M_\pm^2 = \frac{M_W^2}{8} \left( 2 + a_\Phi^2 + b_\Phi^2  + 4\mu^2 \pm 
\sqrt{(2 + a_\Phi^2 + b_\Phi^2  - 4\mu^2)^2 + 16\mu^2 a_\Phi^2} \right)
\label{eqn:evalues-SMXStuk}
\end{equation}
Identifying the lighter of these gauge bosons with the $Z$ and equating 
$M_- = M_Z = M_W/ \cos \theta_W$ leads to the equation
\begin{equation}
\mu^2 \left( b_\Phi^2 - 2 \tan^2\theta_W \right) =  \frac{1}{2\cos^2\theta_W}
\left( a_\Phi^2 + b_\Phi^2 - 2 \tan^2\theta_W \right)
\end{equation}
There are now two possibilities, viz.
\vspace*{-12pt}
\begin{enumerate}
\item We have $a_\Phi$ arbitrary and choose
\begin{equation}
b_\Phi^2 \neq 2 \tan^2 \theta_W \qquad\qquad
\mu^2 = \frac{1}{2 \cos^2\theta_W} \left( 1 + \frac{a_\Phi^2}{b_\Phi^2 
- \tan^2 \theta_W} \right)
\label{eqn:unphysical-SMXStuk}
\end{equation}
\item We have $\mu^2$ arbitrary and choose
\begin{equation}
a_\Phi = 0 \qquad\qquad\qquad
b_\Phi^2 = 2 \tan^2 \theta_W
\label{eqn:physical-SMXStuk}
\end{equation}
\end{enumerate}
\vspace*{-12pt}
Both options look equally plausible at this stage, but it can be shown 
(see Appendix) that the first case (\ref{eqn:unphysical-SMXStuk}) leads 
to an electric charge operator which is different from the SM. As this 
cannot be, we are forced to choose the second option 
(\ref{eqn:physical-SMXStuk}). This simplifies the mass matrix in 
Eq.~(\ref{eqn:matrix-SMXStuk}) considerably, to
\begin{equation}
\mathbb{M} = \frac{M_W^2}{4} \left( \begin{array}{ccc}
2 & 0 & -2 \tan\theta_W \\
0 & 4\mu^2 & 0 \\
-2 \tan\theta_W & 0 & 2 \tan^2 \theta_W \end{array} \right)
\label{eqn:matrix-simple-SMXStuk}
\end{equation} 
which leads to mass terms
\begin{equation}
{\cal L}_{\rm M} = M_W^2 W^{+\mu}W_\mu^- + \frac{1}{2} M_Z^2 Z^\mu Z_\mu 
+ \frac{1}{2} M_{Z'}^2 Z'^\mu Z'_\mu 
\label{eqn:massterms-SMXStuk}
\end{equation}
where
\begin{equation}
M_Z = \frac{M_W}{\cos \theta_W} \qquad\qquad   
M_{Z'} = \sqrt{2} M_W \mu = M \sqrt{2 + \frac{\lambda'_2 v^2}{M^2}} 
\label{eqn:MZ1-SMXStuk}
\end{equation}
This model thus has two extra fields, the scalar $\sigma$ and the gauge 
boson $Z'$, both of whose masses depend on the unknown parameters $M$ 
and $\lambda'_{1,2}$. Inspection of Eqs.~(\ref{eqn:scalarmass-SMXStuck}) 
and (\ref{eqn:MZ1-SMXStuk}) shows that these masses can be made 
arbitrarily large by making $M$ arbitrarily large, which would cause 
these fields to effectively decouple from the SM part of the Lagrangian.

Completing the diagonalisation, the physical states corresponding to 
these gauge bosons are now
\begin{eqnarray}
Z_\mu  & = & W_\mu^3 \cos\theta_W - \widetilde{C}_\mu \sin\theta_W \nonumber \\
A_\mu  & = & W_\mu^3 \sin\theta_W - \widetilde{C}_\mu \cos\theta_W \nonumber \\
Z'_\mu & = & \widetilde{B}_\mu
\label{eqn:gauge-bosons-SMXStuk}
\end{eqnarray}
so that the gauge-fixing in Eq.~(\ref{eqn:fixing-SMXStuck}) is actually 
relevant to the $Z'$ field. This mixing pattern is very close to the SM, 
with the photon and the $Z$ boson being mixed states while the $Z'$ 
states stands apart. Of course, we do have an extra mixing between the 
two $U(1)$ gauge bosons, which happens due to the gauge-fixing terms in 
Eq.~(\ref{eqn:gauge-bilinears-SMXStuck}), irrespective of the 
symmetry-breaking in the Higgs sector.

All that remains now is to rewrite the interaction Lagrangian 
(\ref{eqn:model-SMXStuck}) in terms of these physical gauge bosons and 
derive the Feynman vertices for the model. This is described in the next 
section. Before that, however, it is necessary to introduce the fermions 
i.e. leptons and quarks into this model. Let $\Psi(x)$ be an arbitrary 
$SU(2)$ doublet of fields, with $U(1)$ gauge charges $Y$ and $X$ 
respectively. The covariant derivative acting on this is
\begin{equation}
{\cal D}_\mu = \mathbb{I} \partial_\mu - ig\mathbb{T}_a W_\mu^a 
-\frac{i}{2} g'_Y Y B_\mu \mathbb{I} -\frac{i}{2} g'_X X C_\mu \mathbb{I} 
\label{eqn:covariantgeneral-SMXStuk}
\end{equation}
which can be written in terms of the physical fields as 
\begin{equation}
{\cal D}_\mu = \mathbb{I} \partial_\mu 
- ig\left( \mathbb{T}_+ W_\mu^+ + \mathbb{T}_- W_\mu^- \right)  
- ig\cos\theta_W {\cal Q}_Z Z_\mu 
- ig\sin\theta_W {\cal Q} A_\mu 
- ig{\cal Q}_Z^\prime Z'_\mu 
\label{eqn:generators-SMXStuk}
\end{equation}
where $\mathbb{T}_\pm = \mathbb{T}_1 \pm i\mathbb{T}_2$ and $W_\mu^\pm = \left(W_\mu^1
\mp iW_\mu^2\right)/\sqrt{2}$ as in the SM, and the other generators are
\begin{equation}
{\cal Q}_Z = \mathbb{T}_3 - \frac{b\tan\theta_W}{2\sqrt{2}}\mathbb{I}
\qquad\qquad
{\cal Q} = \mathbb{T}_3 + \frac{b}{2\sqrt{2}\tan\theta_W}\mathbb{I}
\qquad\qquad
{\cal Q}_Z^\prime = \frac{a}{2\sqrt{2}}\mathbb{I}
\label{eqn:Qoperators-SMXStuk}
\end{equation}
in terms of
\begin{equation}
a = \frac{1}{g} \left( g'_Y Y + g'_X X \right)
\qquad\qquad
b = \frac{1}{g} \left( g'_Y Y - g'_X X \right)
\label{eqn:ab-general-SMXStuk}
\end{equation}
For the scalar doublet $\Psi(x) = \Phi(x)$, we have $a = a_\Phi = 0$ and 
$b = b_\Phi = \sqrt{2} \tan\theta_W$, i.e.  
\begin{equation}
g'_Y Y_\Phi = - g'_X X_\Phi = \frac{g \tan\theta_W}{\sqrt{2}} = \frac{g'}{\sqrt{2}}
\label{eqn:gXgY-scalar-relation-SMXStuk}
\end{equation}
where $g'$ is the $U(1)_Y$ coupling of the SM. The generators become
\begin{equation}
\mathbb{Q}_Z^{(\Phi)} = \mathbb{T}_3 - \frac{1}{2} \tan^2\theta_W \mathbb{I}
\qquad\qquad
\mathbb{Q}^{(\Phi)} = \mathbb{T}_3 + \frac{1}{2}\mathbb{I}
\qquad\qquad
\mathbb{Q}_{Z^\prime}^{(\Phi)} = 0
\label{eqn:generators-final-SMXStuk}
\end{equation}
In general, the generators ${\cal Q}_Z$ and ${\cal Q}$ are identical 
with the SM and we can identify $e = g\sin\theta_W$, exactly as in the 
SM. This shows that the interactions of the $Z$ and the photon are the 
same as in the SM. Since $\mathbb{Q}_{Z^\prime}^{(\Phi)} = 0$, the $Z'$ 
field decouples from the Higgs doublet. However, we do not have enough 
constraints to determine $g'_Y$, $g'_X$, $Y_\Phi$ and $X_\Phi$ uniquely, 
and therefore, to proceed further, we must rely on some ans\"{a}tze. We, 
therefore, treat the $g'_Y$ and $g'_X$ as free parameters, which 
immediately leads to the scalar having quantum numbers
\begin{equation}
Y_\Phi = \frac{g'}{\sqrt{2} g'_Y}  \qquad\qquad  X_\Phi = - \frac{g'}{\sqrt{2} g'_X}
\label{eqn:XY-scalars-SMXStuk}
\end{equation} 
In the general case, the values of $a_\Phi$ and $b_\Phi$ also enable us 
to write Eq.~(\ref{eqn:ab-general-SMXStuk}) as
\begin{equation}
a = \frac{g'}{\sqrt{2}g} \left( \frac{Y}{Y_\Phi} - \frac{X}{X_\Phi} \right)
\qquad\qquad
b = \frac{g'}{\sqrt{2}g} \left( \frac{Y}{Y_\Phi} + \frac{X}{X_\Phi} \right)
\label{eqn:ab-general-scalar-SMXStuk}
\end{equation} 
We must identify the charge operator ${\cal Q}$ with that of the SM, 
which immediately leads to
\begin{equation}
b = \sqrt{2} \tan\theta_W Y_{\rm SM}
\end{equation} 
which reduces to
\begin{equation}
\frac{Y}{Y_\Phi} + \frac{X}{X_\Phi} = 2Y_{\rm SM}
\label{eqn:XY-constraintl-SMXStuk}
\end{equation}
where $Y_{SM}$ is the weak hypercharge normally assigned to the gauge 
multiplet in the SM. Eq.~(\ref{eqn:XY-constraintl-SMXStuk}) is obviously 
consistent with the SM value $Y_{\rm SM}^\Phi = +1$. Substitution for 
$Y_\Phi$ and $X_\Phi$ using Eq.~(\ref{eqn:XY-scalars-SMXStuk}) leads to
\begin{equation}
X = -\frac{\sqrt{2}g'}{g_X} Y + \frac{g_Y}{g_X} Y_{\rm SM}
\label{eqn:X-value-SMXStuk}
\end{equation}
and therefore
\begin{equation}
a = \frac{2g_Y}{g} Y - \frac{\sqrt{2}g'}{g} Y_{\rm SM} 
\label{eqn:a-value-SMXStuk}
\end{equation}
For fermions, we must choose the $Y$ and $X$ quantum numbers such than 
there is no chiral anomaly in the theory, which leads to a set of 
anomaly cancellation conditions, which can be written in the usual 
shorthand as
\begin{equation}
{\rm Tr}[\mathbb{Y}] = {\rm Tr}[\mathbb{X}] = 0 
\qquad\qquad
{\rm Tr}[\mathbb{Y}^3] = {\rm Tr}[\mathbb{Y}^2 \mathbb{X}] 
= {\rm Tr}[\mathbb{Y} \mathbb{X}^2] = {\rm Tr}[\mathbb{X}^3] = 0 
\label{eqn:anom-cancel-SMXStuk}
\end{equation}
For the SM, it is well known that ${\rm Tr}[\mathbb{Y}_{\rm SM}] = {\rm 
Tr}[\mathbb{Y}_{\rm SM}^3] = 0$, and hence the simplest ansatz we can 
take for the hypercharges in this model is to choose $Y \propto Y_{\rm 
SM}$, which, by Eq.~(\ref{eqn:X-value-SMXStuk}) implies $X \propto 
Y_{\rm SM}$ as well. Making this choice, we can now rewrite 
Eq.~(\ref{eqn:a-value-SMXStuk}) as
\begin{equation}
a = \eta Y_{\rm SM}
\end{equation}
where $\eta$ is a free parameter which is the same for all fermions. 
This will show up in all the couplings of the $Z'$ boson.

We thus have a model with six free parameters, viz. the mass scale $M$ 
(which can be exchanged for the physical mass of the $Z'$ boson), the 
two gauge couplings $g'_X$ and $g'_Y$, the two quartic couplings 
$\lambda'_1$ and $\lambda'_2$, and finally the $\eta$ parameter. The 
presence of two extra fields -- the scalar $\sigma$ and the gauge boson 
$Z'$ -- will lead to the existence of some extra interaction vertices in 
this model. These are worked out in the next section.

\section{New interactions}
\setcounter{equation}{0}

\begin{center} 
\begin{figure}[h!] 
\centerline{\includegraphics[width=0.6\textwidth]{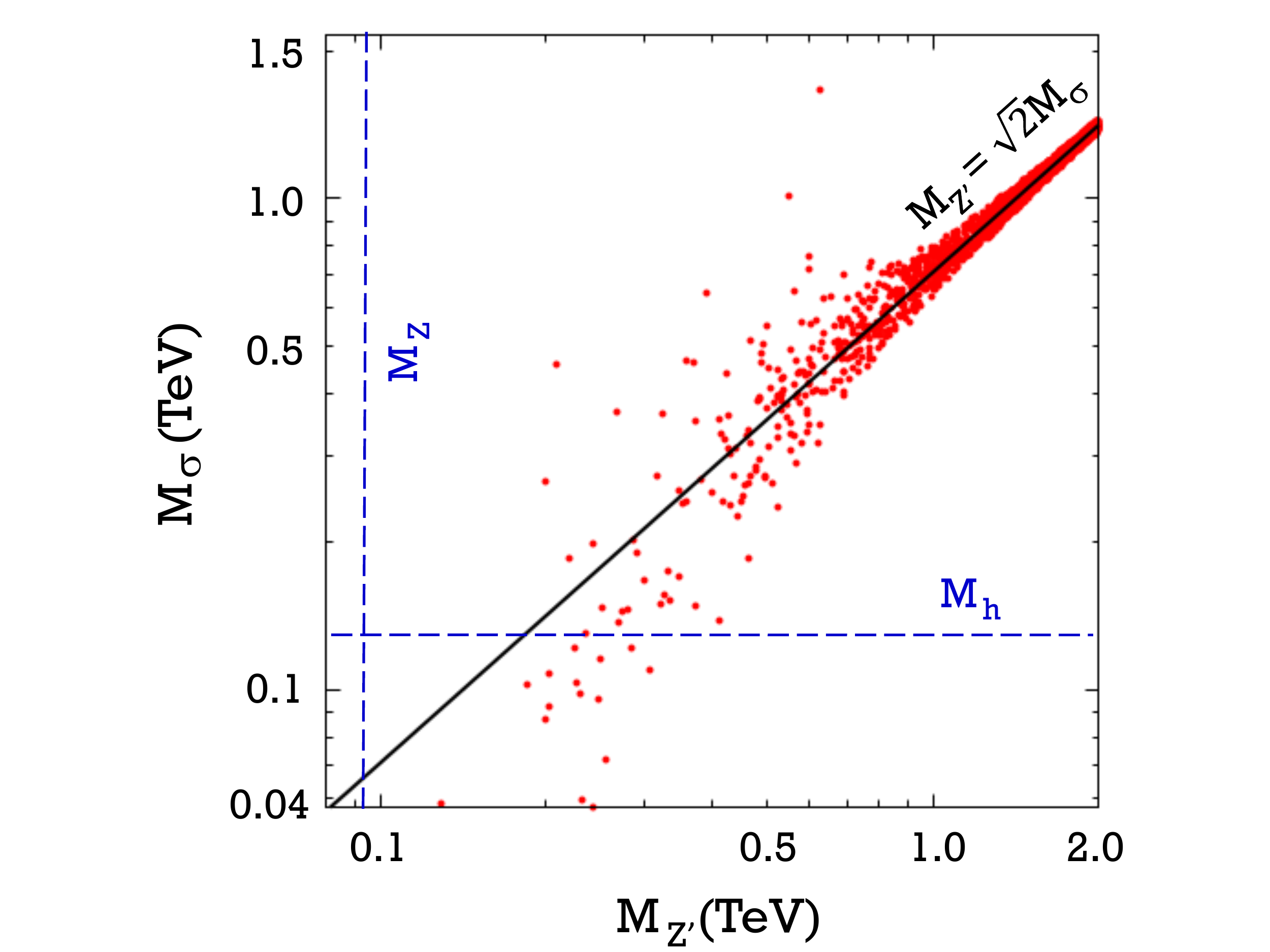} } 
\vspace*{-0.1in} 
\caption{\setstretch{1.05}\footnotesize Masses of the St\"{u}ckelberg 
scalar $\sigma$ and the extra gauge boson $Z'$ when $\zeta'$, 
$\lambda'_1$ and $\lambda'_2$ are varied as follows: $0.137 \leq \zeta' 
\leq 100$ and $-\sqrt{4\pi} \leq \lambda'_{1.2} \leq +\sqrt{4\pi}$. For 
heavier masses ($M_{Z'} > 1$~TeV), it may be seen that the ratio 
$M_{Z'}/M_\sigma \to \sqrt{2}$. The masses $M_Z$ of the $Z$ boson and $M_h$ of 
the Higgs boson are marked on the plot for comparison.}
\label{fig:spectrum} 
\end{figure} 
\end{center}

We have, till now, considered only the bilinear and mass terms in the 
Lagrangian of Eq.~(\ref{eqn:model-SMXStuck}). As elsewhere, there are 
also cubic and quartic terms in the interaction Lagrangian, which will 
give rise to the vertices of the theory in the purely bosonic sector. In 
this section, we evaluate them and work out the corresponding Feynman 
rules. However, the first step towards this is write the new mass 
parameters, using Eqs.~(\ref{eqn:scalarmass-SMXStuck}) and 
(\ref{eqn:MZ1-SMXStuk}) as
\begin{equation}
M^2 = \frac{1}{2}v^2 \left(\zeta' - \lambda'_2 \right)
\qquad\qquad
M_\sigma^2 = \frac{1}{2}v^2 \frac{\zeta'\left(\zeta' - \lambda'_2 \right)}
{\zeta' + \lambda'_1 - \lambda'_2}
\end{equation}
where $v$ is the SM Higgs vev and $\zeta' = M_{Z'}^2/v^2$. If we assume 
the mass of the $Z'$ boson to range from $M_Z$ to about 2.5~TeV, then 
the range of $\zeta'$ will be from $0.137$ to $100$. The values of 
$\lambda'_1$ and $\lambda'_2$ are not constrained except by the 
corresponding perturbative limits $\pm\sqrt{4\pi}$. The corresponding 
values of $M_\sigma$ and $M_{Z'}$ are plotted in 
Figure~\ref{fig:spectrum}. We do not plot $M$ as it is not a physical 
mass.

\vspace*{-0.2in} 
\vspace*{-0.5in}

\begin{center} 
\begin{figure}[h!] 
\centerline{\includegraphics[width=0.9\textwidth]{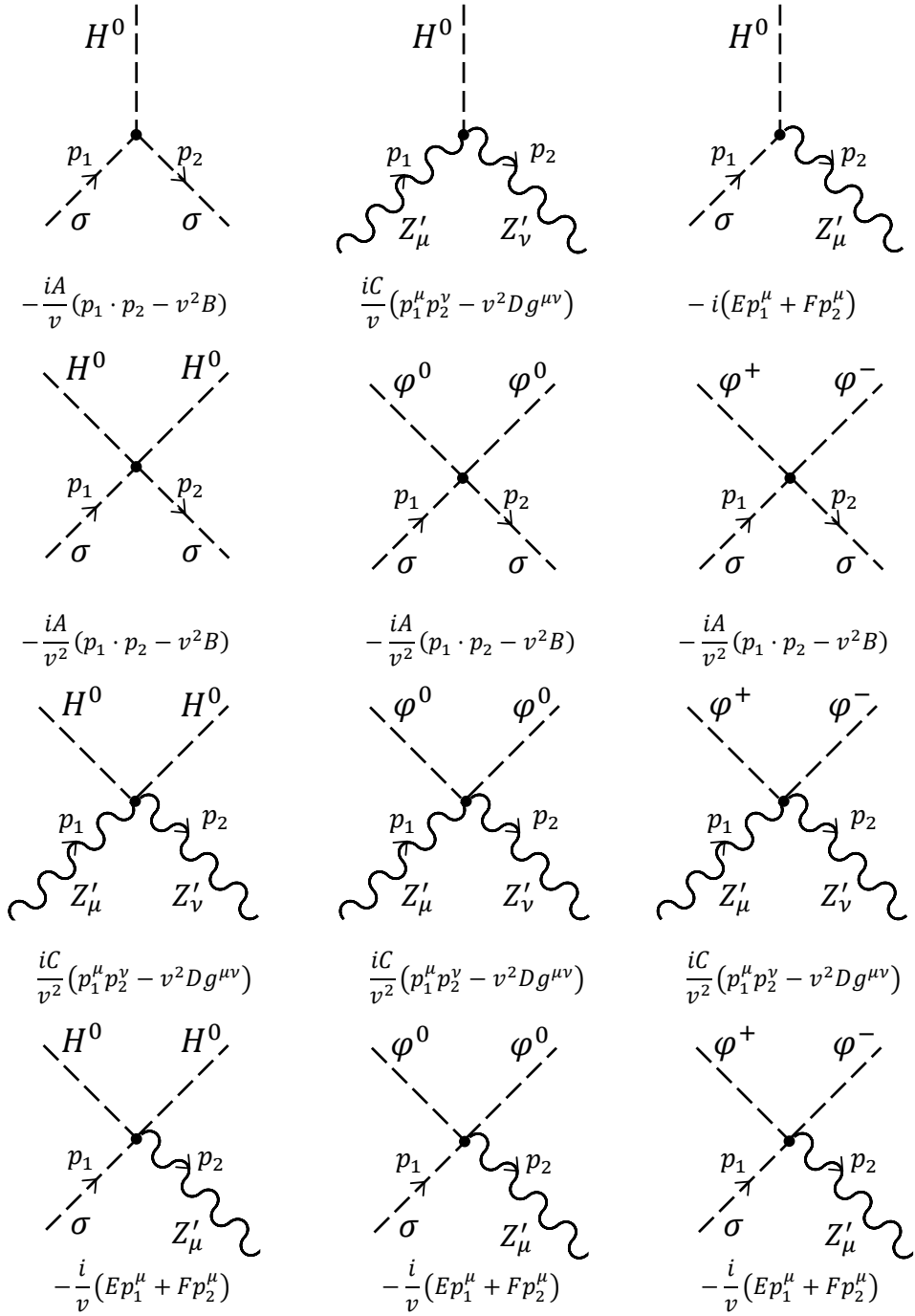} } 
\vspace*{-0.4in} 
\caption{\setstretch{1.05}\footnotesize Feynman vertices for the new bosonic interactions
of the $\sigma$ and $Z'$ fields.}  
\label{fig:Feynmanvertices} 
\end{figure} 
\end{center}

Though we do not make a detailed analysis in this work, it is likely 
that the lower end of the spectrum, especially the scalar masses lower 
than 100~GeV, may be ruled out by the existing data. However, there will 
be plenty of parameter choices which will easily evade these bounds, 
since the scalar and the $Z'$ will simultaneously become heavy. In any 
case, to make a detailed phenomenological analysis, the Feynman rules 
for the model require to be worked out in detail. In this model, when we 
work out the cubic and quartic interactions of the model of 
Eq.~(\ref{eqn:model-SMXStuck}), it turns out that the interactions of 
the photon and the $Z$-boson are identical with those of the SM. Thus, 
one cannot search for signatures of this model by looking for deviations 
in the usual SM signatures. However, the $Z'$ boson and the scalar 
$\sigma$ will have some new interactions as in the interaction 
Lagrangian below.
\begin{eqnarray}
{\cal L}_{\rm int} = \frac{\mid\!\Phi\!\mid^2}{v^2} \left[
A\left(\partial^\mu \sigma \; \partial_\mu \sigma - Bv^2 \sigma^2 \right)
- C\left\{ \left(\partial^\mu Z'_\mu\right)^2 - Dv^2 Z^{\prime\mu} Z'_\mu \right\}
+ v\left( E Z'_\mu \partial^\mu\sigma + F \sigma \partial^\mu Z'_\mu \right) \right]
\nonumber \\
\end{eqnarray}
where
\begin{equation}
\mid\!\Phi\!\mid^2 = \varphi^+\varphi^- + \frac{1}{2} \left(h^0\right)^2 
+ \frac{1}{2} \left(\varphi^0\right)^2 + vh^0 
\end{equation}
and
\begin{eqnarray}
A &=& \frac{2\lambda'_1}{\zeta' + \lambda'_1 - \lambda'_2} 
\nonumber \\ [2mm]
B &=& \frac{\lambda'_2(\zeta' - \lambda'_2)}{2\lambda'_1} 
\nonumber \\ [2mm]
C &=& \frac{4\lambda'_2(\zeta' + \lambda'_1 - \lambda'_2)}{\zeta'(\zeta' - \lambda'_2)}
\nonumber \\ [2mm]
D &=& \frac{\zeta^{\prime 2}\lambda'_1(\zeta' - \lambda'_2)}
{2\lambda'_2(\zeta' + \lambda'_1 - \lambda'_2)^2} 
\nonumber \\ [2mm]
E &=& \frac{2\lambda'_1 \sqrt{\zeta'}}{\zeta' + \lambda'_1 - \lambda'_2}
\nonumber \\ [2mm]
F &=& \frac{2\lambda'_2}{\sqrt{\zeta'}}
\label{eqn:ABCDEF}
\end{eqnarray}
The vertices of these interactions can be read off from this equation, 
and are listed in Figure~\ref{fig:Feynmanvertices}. As a quick check, we 
can see that in the decoupling limit, i.e. when $\zeta' \to \infty$ 
(equivalently, when $\lambda'_{1,2} \to 0$), the parameters in 
Eq.~(\ref{eqn:ABCDEF}) reduce to
\begin{eqnarray}
A &\to& \frac{2\lambda'_1}{\zeta'} \qquad\qquad\qquad 
C \to \frac{4\lambda'_2}{\zeta'} \qquad\qquad\qquad 
E \to \frac{2\lambda'_1}{\sqrt{\zeta'}} \nonumber \\ [2mm] 
B &\to& \frac{\zeta' \lambda'_2}{2\lambda'_1} \ \ \qquad\qquad\quad
D \to \frac{\zeta'\lambda'_1}{2\lambda'_2} \qquad\qquad\quad \ \ 
F \to \frac{2\lambda'_2}{\sqrt{\zeta'}} 
\end{eqnarray}

\vspace*{-0.5in}

Thus the only couplings which do not vanish are
\begin{equation}
AB \to \lambda'_2 \qquad\qquad\qquad  CD \to 2\lambda'_1
\end{equation} 
which remain perturbative because the $\lambda'_{1,2}$ are perturbative. 
However, these may get further constrained by the non-observation of the 
interactions of Figure~\ref{fig:Feynmanvertices}. A related issue is the 
presence of derivative couplings and hence a strong momentum-dependence 
of the cubic and quartic couplings, which could, in principle, lead to 
unitarity violation at a scale significantly higher than the electroweak 
scale. This does not seem to happen in $U(1)$ St\"{u}ckelberg models 
\cite{Kuzmin}. However, such an analysis for the present model could be 
interesting in its own right, for, if unitarity is violated, it would 
mean that the St\"{u}ckelberg mechanism only works within the framework 
of an effective theory. However, we defer this analysis to a future work 
\cite{selves}.

\vspace*{-0.2in} 
\begin{center} 
\begin{figure}[h!] 
\centerline{\includegraphics[width=0.65\textwidth]{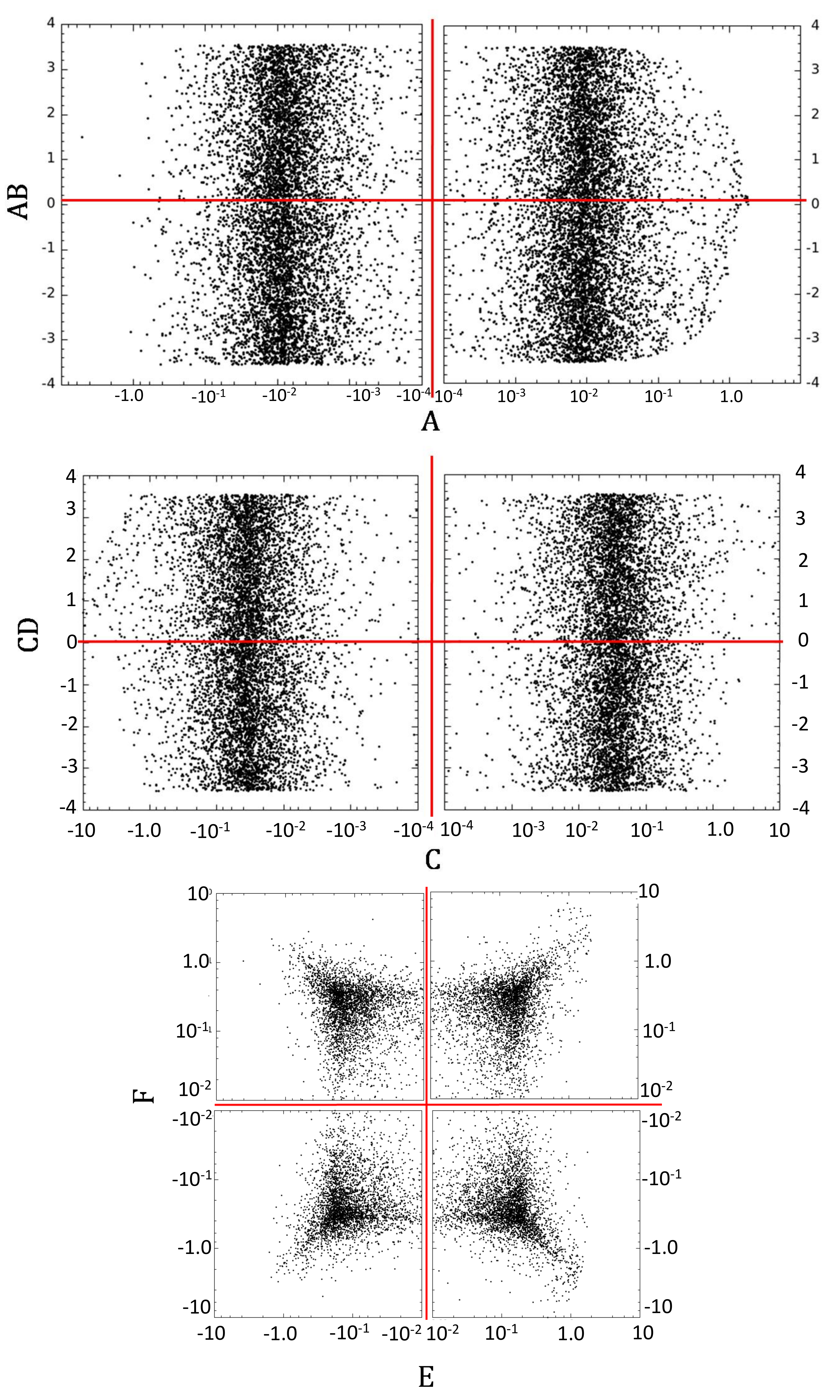} } 
\vspace*{-0.1in} 
\caption{\setstretch{1.05}\footnotesize Scatter plots showing the 
variation of the coupling parameters $A$, $AB$, $C$, $CD$, $E$ and $F$ 
with random variations in the parameters $\zeta'$, $\lambda'_1$ and 
$\lambda'_2$ within their allowed ranges. The coordinate axes are
marked in red.} \label{fig:scatterplots}
\end{figure} 
\end{center} 
\vspace*{-0.5in} 

Nevertheless, even apart from the decoupling limit, we can form an idea 
about the strength of the couplings $A$, $AB$, $C$, $CD$, $E$ and $F$ by 
varying the independent parameters $\zeta'$, $\lambda'_1$ and 
$\lambda'_2$ between their allowed ranges, as we have done above for the 
masses of the $\sigma$ and the $Z'$. The results are shown in 
Figure~\ref{fig:scatterplots}. It may be seen that the parameters $A$ 
and $C$ are mostly restricted to the order $10^{-3}$ and $10^{-2}$ 
respectively, while the product couplings $AB$ and $CD$ vary almost 
uniformly over the full allowed range. The couplings $E$ and $F$ remain 
mostly smaller than unity, with a few outlying values. We may conclude, 
therefore, that even if perturbative unitarity is breached in this 
model, it will happen at a energy at least an order of magnitude above 
the electroweak scale, i.e. a few TeV, which is currently unattainable 
at existing collider machines.

\begin{center} 
\begin{figure}[h!] 
\centerline{\includegraphics[width=0.45\textwidth]{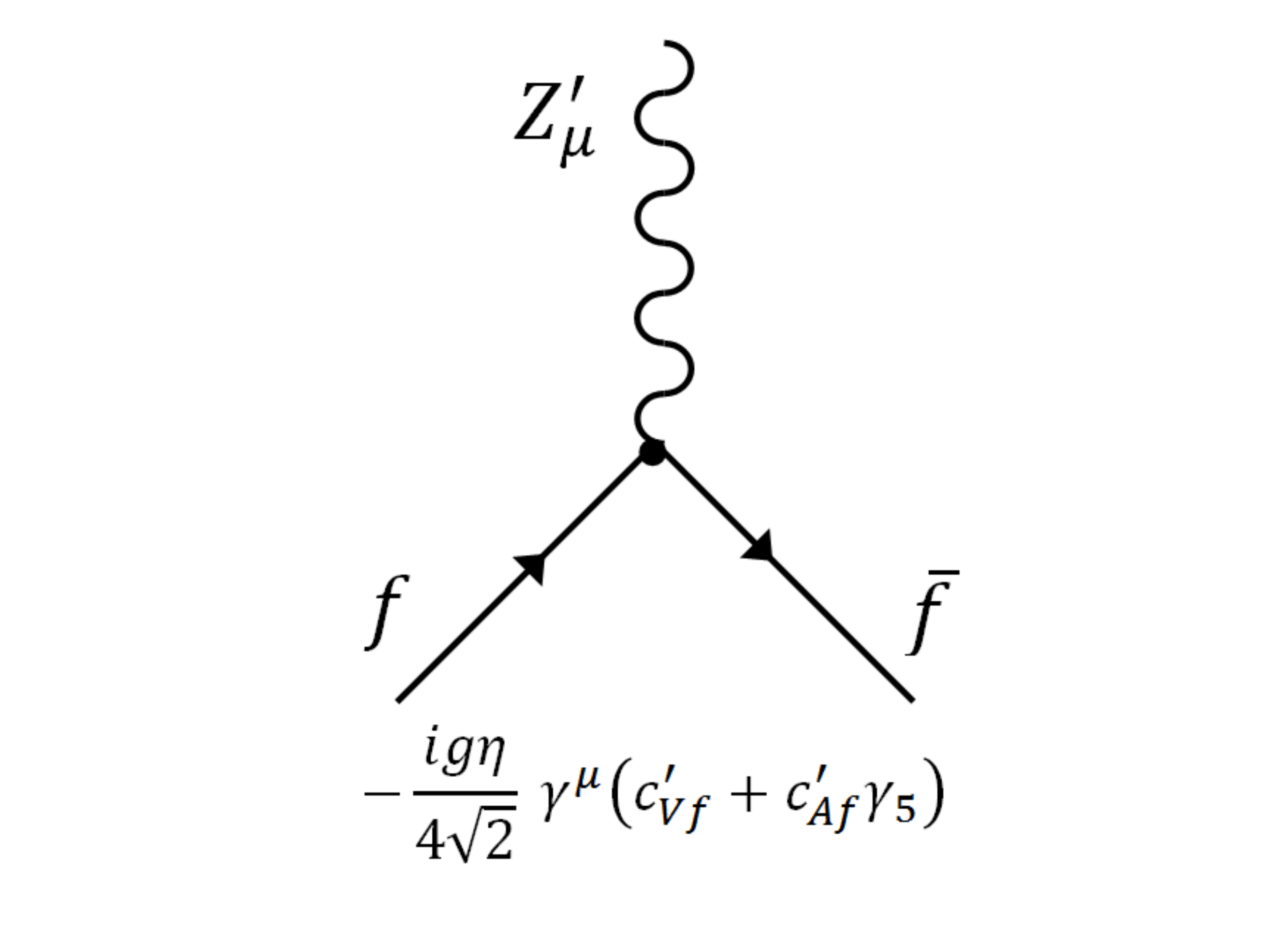} } 
\vspace*{-0.2in} 
\caption{\setstretch{1.05}\footnotesize Feynman vertex for the $Z'$ 
coupling with fermions. The constants $c'_{Vf}$ and $c'_{Af}$ are listed 
in Table~\ref{tab:fermion-couplings}. } \label{fig:fermionvertices}
\end{figure} 
\end{center}

Finally, we come to the gauge-fermion interactions. As usual, these 
arise from
\begin{eqnarray}
{\cal L}_f = &\sum_{i=k}^3& \bigl[ 
i\bar{L}_{Lk} \gamma^\mu {\cal D}_\mu^{(L)} L_{Lk} 
+ i \bar{\ell}_{Rk} \gamma^\mu D_\mu^{(\ell)} \ell_{Rk}  \nonumber \\ [2mm]
&& \!\!\!\!\!  + \ i\bar{Q}_{Lk} \gamma^\mu {\cal D}_\mu^{(Q)} Q_{Lk}  
+ i \bar{u}_{Rk} \gamma^\mu D_\mu^{(u)} u_{Rk} 
+ i \bar{d}_{Rk} \gamma^\mu D_\mu^{(d)} d_{Rk} 
\bigr]
\end{eqnarray}
where the covariant derivatives ${\cal D}_\mu$ on the $SU(2)_L$ doublets 
$L_L^T = (\!\!\begin{array}{cc} \nu_{L\ell} & \ell_L \end{array}\!\!)$ 
and $Q_L^T = (\!\!\begin{array}{cc} u_{L} & d_L \end{array}\!\!)$ are 
defined in Eq.~(\ref{eqn:generators-SMXStuk}) and the covariant 
derivative on the singlets $\ell_R$, $u_R$ and $d_R$ are defined as
\begin{equation}
D_\mu = \partial_\mu 
- ig\cos\theta_W Q_Z Z_\mu 
- ig\sin\theta_W Q A_\mu 
- ig Q_Z^\prime Z'_\mu 
\label{eqn:generators-singlet-SMXStuk}
\end{equation}
where
\begin{eqnarray}
Q_Z &=&  - \frac{b\tan\theta_W}{2\sqrt{2}}
= \frac{1}{2} Y_{\rm SM} \tan^2\theta_W  
\nonumber \\
Q &=& \frac{b}{2\sqrt{2}\tan\theta_W}
= \frac{1}{2} Y_{\rm SM} 
\nonumber \\
Q_{Z^\prime} &=& \frac{a}{2\sqrt{2}} = \frac{\eta}{2\sqrt{2}} Y_{\rm SM}
\label{eqn:Qoperators-singlet-SMXStuk}
\end{eqnarray}
In Eq.~(\ref{eqn:Qoperators-singlet-SMXStuk}), the $Z$ boson and photon 
interactions are the same as in the SM, as was the case with the Higgs 
and SM gauge sectors. Thus we will get new interactions only from the 
$Z'$ terms, which can be worked out as
\begin{equation}
{\cal L}_{Z'f\bar{f}} = 
\sum_f -\frac{g\eta}{4\sqrt{2}} \overline{\psi}_f \gamma^\mu 
\left( c'_{Vf} + c'_{Af} \gamma_5 \right) \psi_f
\end{equation}
where the generation-independent constants $c'_{Vf}$ and $c'_{Af}$ are 
listed in Table~\ref{tab:fermion-couplings}. This leads to the Feynman 
vertices in Figure~\ref{fig:fermionvertices}.

It must be noted that this pattern of couplings is not unique, but 
arises only when the simplest ansatz for anomaly cancellation, i.e. $Y 
\propto Y_{\rm SM}$ is taken. Naturally, this retains the 
generation-universality observed in the SM and may not be the best 
choice to explain the flavour anomalies mentioned in the Introduction. 
However, in that case, one can always pick up one or other of the many 
different ans\"{a}tze proposed in $U(1)$-extended SM scenarios 
\cite{Allanach}, and use it in conjunction with the mass generation 
mechanism proposed in this work.

\vspace*{-0.2in} 
\vspace*{-0.5in} 

\begin{center} 
\begin{figure}[h!] 
\centerline{\includegraphics[width=1.1\textwidth]{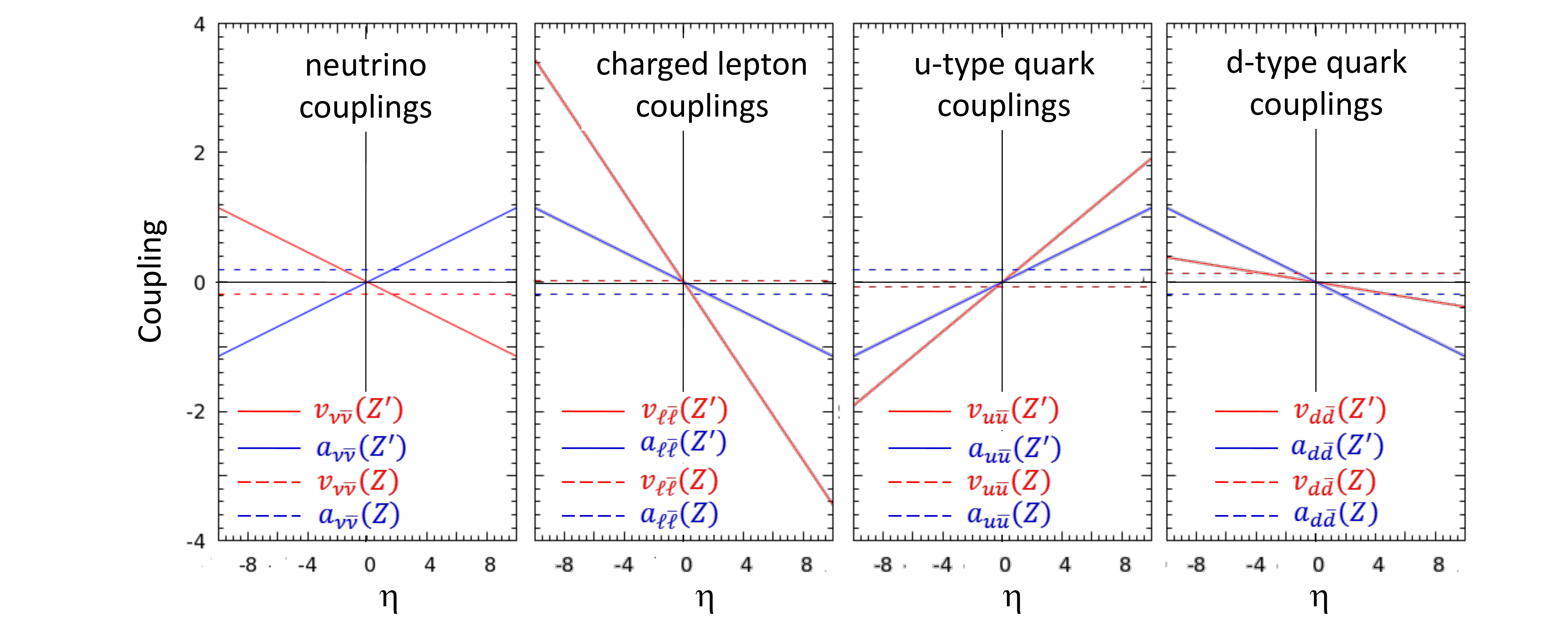} } 
\vspace*{-0.2in} 
\caption{\setstretch{1.05}\footnotesize Fermion couplings of the $Z$ and 
$Z'$ bosons, showing their variation with the parameter $\eta$. The 
range $-10 \leq \eta \leq 10$ ensures that all these couplings stay 
within the perturbative limit.}
\label{fig:fermioncouplings} 
\end{figure} 
\end{center}

To get a comparison between the couplings of the $Z'$ boson with those 
of the $Z$ boson, we have plotted the vector and axial vector couplings 
of both in Figure~\ref{fig:fermioncouplings}. The quantities plotted for 
the $Z$ boson are
\begin{equation}
v_{f\bar{f}}(Z) = \frac{g}{4\cos\theta_W} c_{Vf} \qquad\qquad
a_{f\bar{f}}(Z) = \frac{g}{4\cos\theta_W} c_{Af}
\end{equation}
and for the $Z'$ boson are
\begin{equation}
v_{f\bar{f}}(Z') = \frac{g\eta}{4\sqrt{2}} c'_{Vf} \qquad\qquad
a_{f\bar{f}}(Z') = \frac{g\eta}{4\sqrt{2}} c'_{Af}
\end{equation}
where, in the four panels, from left to right, $f = \nu, \ell, u$ and 
$d$ respectively.

\begin{table}[b!]
\begin{center}
\begin{tabular}{|r||r|r|}
\hline  & & \\ [-2mm]
$f$ & 
$c'_{Vf}$ \hspace*{0.3in} & 
$c'_{Af}$ \hspace*{0.3in} \\ [2mm] \hline\hline  & & \\ [-2mm]
$\nu_\ell$ & 
$Y_{\rm SM}^{(L)} = 1$ &  
$-Y_{\rm SM}^{(L)} = -1$ \\ [2mm] \hline  & & \\ [-2mm]
$\ell$ & 
$Y_{\rm SM}^{(L)} + Y_{\rm SM}^{(\ell)} = 3$ & 
$-Y_{\rm SM}^{(L)} + Y_{\rm SM}^{(\ell)} = 1$ \\ [2mm] \hline  & & \\ [-2mm]
$u$ & 
$Y_{\rm SM}^{(Q)} + Y_{\rm SM}^{(u)} = -\frac{5}{3}$ &
$-Y_{\rm SM}^{(Q)} + Y_{\rm SM}^{(u)} = -1$ \\ [2mm] \hline  & & \\ [-2mm]
$d$ & 
$Y_{\rm SM}^{(Q)} + Y_{\rm SM}^{(d)} = \frac{1}{3}$ &
$-Y_{\rm SM}^{(Q)} + Y_{\rm SM}^{(d)} = 1$ \\ [2mm] \hline
\end{tabular}
\caption{\footnotesize Couplings of the fermions with the $Z'$ boson 
assuming that all the hypercharges are proportional to those in the SM.}
\label{tab:fermion-couplings}
\end{center}
\end{table}

A glance at the figure immediately shows that the couplings of the $Z'$ 
boson to fermions can be significantly stronger than those of the $Z$ 
boson, and hence, in regions which are kinematically favourable, 
production cross-sections of the $Z'$ boson as resonances in fermion 
pair annihilation can be significantly higher than those of the $Z$ 
boson. Of course, we also have the decoupling limit $\eta \to 0$ in 
which the $Z'$ has no fermion interactions whatsoever. This, however, is 
unlikely, since it would call for a fine tuning $Y = g'Y_{\rm 
SM}/\sqrt{2}$ for every fermion hypercharge. We may thus conclude that 
the $Z'$ can have interesting signals at high energy colliders such as 
the LHC and its upgrades, no less than at a high energy $e^+e^-$ 
collider. The study of these will be taken up in a future work 
\cite{selves}.

\vspace*{-0.2in} 
\vspace*{-0.5in}

\section{Concluding remarks}
\setcounter{equation}{0}

In this article, we have reported a careful and detailed development of 
a $U(1)_X$ extended SM, with a St\"{u}ckelberg mechanism to generate a 
mass for the extra $Z'$ gauge boson instead of a further extension of 
the Higgs sector. Our analysis differs from previous studies in that we 
have made the cancellation of (unphysical) bilinear terms in the 
Lagrangian a keystone of our analysis. As a result, we have obtained 
constraints which not only provide a gauge-fixing term for the extra 
gauge boson, but also render the mass matrix extremely simple. This has 
made it possible to consider, for the first time, new quartic 
interactions of the St\"{u}ckelberg scalar, which are permitted by the 
gauge symmetry. Even with these new interactions, however, the model 
remains quite minimal and is very economical in new fields and 
parameters. There are just two new particles, viz., a vector $Z'$ and a 
St\"{u}ckelberg scalar $\sigma$. Similarly, there are just three 
parameters in the bosonic part of the theory, which we have chosen to be 
the $Z'$ mass (scaled to the SM vev) and the two new quartic couplings. 
The advantage of using coupling constants as parameters is, of course, 
their phenomenological limitation within the perturbative range 
$-\sqrt{4\pi} < \lambda'_{1,2} < +\sqrt{4\pi}$. Thus, the mass scale for 
new physics is entirely set by the $Z'$ mass.

Another nice feature of the model presented here is that the requirement 
that we obtain the correct charge operator on the Higgs doublet acts as 
a second constraint, which renders the mass matrix even more simple. As 
a result, the mixing pattern of gauge bosons becomes such that the 
interactions of the photon and the $Z$ boson become identical with those 
in the SM. This immediately renders the model immune from constraints 
arising from measurement of $Z$ boson interactions, but alas! it also 
removes a possible phenomenological handle on the new physics. 
Nevertheless, there do exist a set of interaction vertices between the 
physical and unphysical fields in the Higgs doublet with the $Z'$ and 
the $\sigma$, which we plan to investigate for possible signals. These 
new vertices, as we have shown, assume small, but not too small values, 
which may make these interactions visible at a higher energy and/or 
luminosity machine.

The fermion sector of the model differs from most other $U(1)_X$ 
extended models in that the couplings of the $Z$ boson remain identical 
to those in the SM, with no effect of mixing being manifest. The $Z'$ 
boson also couples to fermions, but these couplings depend only on the 
hypercharge assignments in the model. The requirement that the photon 
couplings match with QED and the SM, and that the chiral anomalies 
should cancel put very stringent constraints on the hypercharge choices. 
This is no different from generic $U(1)_X$- extended models, and the 
same kind of choices are possible. The simplest ansatz, which we have 
adopted in this article, is to take the hypercharges all proportional to 
their SM values. This automatically ensures anomaly cancellation, but it 
drastically reduces the freedom to vary the $Z'$ couplings to fermions. 
While this has the advantage of economy -- the parameter space being 
restricted to that one proportionality factor -- it may prove inadequate 
to explain flavour anomalies as they currently stand. In such a case, 
perhaps a different ansatz for the fermion-$Z'$ couplings may have to be 
adopted. However, a phenomenological analysis would be needed before we 
come to any such conclusion.

To make a long story short, then, we have re-examined the possibility 
that a msssive vector boson $Z'$ in a $U(1)$-extended SM can get its 
mass through a St\"{u}ckelberg mechanism rather than by invoking a 
further complication of the symmetry-breaking mechanism of the SM. The 
answer to this question seems to be in the affirmative. The resultant 
model, which has several features not considered earlier in the existing 
literature, cannot be falsified by current data, but it can predict 
interesting novel signatures at future colliders. It also has the 
potential to explain flavour physics anomalies. A joint study of such 
anomalies and possible collider signals in the context of this model 
will be required in order to make a more definitive statement, and is in 
the pipeline.

\vfill

\hrule
{\sl Acknowledgments}: RV is grateful to A.~Misra for discussions and 
acknowledges the DST-INSPIRE for financial support. SR acknowledges 
extensive discussions with A.~Venkata (CEBS, Mumbai) which helped 
initiate this study. SR acknowledges support of the Department of Atomic 
Energy, Government of India, under Project Identification No. RTI 4002.

\newpage
\setstretch{1.25}

\newpage
\centerline{\Large\bf Appendix}
\renewcommand{\theequation}{A.\arabic{equation}}
\setcounter{equation}{0}
\setstretch{1.15}

In this Appendix, we present the proof that the choice \#~1 in 
Section~4, which is described in Eq.~(\ref{eqn:unphysical-SMXStuk}), 
does not lead to a physically viable solution. For this choice, we note 
that $a_\Phi$ and $b_\Phi$ are arbitrary, and hence $\mu^2$ is fixed by 
Eq.~(\ref{eqn:unphysical-SMXStuk}) as
\begin{equation}
\mu^2 = \frac{1}{2 \cos^2\theta_W} \left( 1 + \frac{a_\Phi^2}{b_\Phi^2 - \tan^2 \theta_W} \right)
\end{equation}
and the mass matrix is given by Eq.~(\ref{eqn:matrix-SMXStuk}) as
\begin{equation}
\mathbb{M} = \frac{M_W^2}{4} \left( \begin{array}{ccc}
2 & -\sqrt{2} a_\Phi & -\sqrt{2} b_\Phi \\
-\sqrt{2} a_\Phi & a_\Phi^2 + 4\mu^2 & a_\Phi b_\Phi \\
-\sqrt{2} b_\Phi & a_\Phi b_\Phi & b_\Phi^2 \end{array} \right)
\end{equation}
which has eigenvalues $M_Z^2 = M_W^2/\cos^2\theta_W$, $M_\gamma^2 = 0$ and
\begin{equation}
M_{Z'}^2 = \frac{M_W^2}{4} \left( 2 + a_\Phi^2 + b_\Phi^2 + 4 \mu^2 +
\sqrt{(2 + a_\Phi^2 + b_\Phi^2 - 4 \mu^2)^2 + 16\mu^2 a_\Phi^2} \right)
\end{equation} 
We can now work out the eigenvectors and obtain the diagonalising matrix 
$\mathbb{S}$ in terms of which we can write
\begin{equation}
\left(\!\! \begin{array}{c} Z_\mu \\ A_\mu \\ Z'_\mu \end{array} \!\!\right) = \mathbb{S}
\left(\!\! \begin{array}{c} W_\mu^3 \\ \widetilde{B}_\mu \\ \widetilde{C}_\mu \end{array} \!\!\right)
\end{equation}
The only relevant elements of $\mathbb{S}$ are
\begin{equation}
S_{12} = S_{32} = \frac{b_\Phi}{\sqrt{2+b_\Phi^2}} \qquad\qquad S_{22} = 0
\end{equation}
and these can be obtained simply from the eigenvector corresponding to 
the vanishing eigenvalue. The covariant derivative of 
Eq.~(\ref{eqn:generators-SMXStuk}) can then be written out in full, but 
these three elements are the only ones needed to obtain the electric 
charge operator
\begin{equation}
ie{\cal Q} = \frac{ig b_\Phi}{\sqrt{2+b_\Phi^2}} 
\left(\mathbb{T}_3 + \frac{b}{b_\Phi}\mathbb{I}   \right)
\end{equation}
To obtain the correct value of $e$ we will have to identify
\begin{equation}
\frac{b_\Phi}{\sqrt{2+b_\Phi^2}} = \sin\theta_W
\end{equation}
which immediately leads to $b_\Phi^2 = 2 \tan^2\theta_W$ contradicting 
Eq.~(\ref{eqn:unphysical-SMXStuk}), our starting point. This completes 
the proof that there is no way we can get the correct electric charge 
quantum $e$ if we choose the option \#~1 in Section 4.

\end{document}